\documentclass[%
 reprint,
 amsmath,amssymb,
 aps,
showkeys
]{revtex4-2}

\usepackage{graphicx}
\usepackage{dcolumn}
\usepackage{bm}

\begin{document}

\preprint{APS/123-QED}

\title{Linearized Gravity in the Starobinsky Model: Perturbative Deviations from General Relativity}

\author{Roger Anderson Hurtado}
\email{rahurtadom@unal.edu.co}
\affiliation{%
 Observatorio Astronómico Nacional, Universidad Nacional de Colombia
}%



\date{\today}

\begin{abstract}
In this work, we linearize the field equations of $f(R)$ gravity using the Starobinsky model, $R+R^2/(6m^2)$, and examine the modifications to General Relativity. We derive an equation for the trace, $T$, of the energy-momentum tensor, which we then decompose using an auxiliary field. This field satisfies the wave equation with $T$ as its source, while simultaneously acting as an effective source for the classical deviation, $\bar h$, governed by the Klein-Gordon equation. The fields were expressed in terms of Green's functions, whose symmetry properties facilitated the solution of the trace equation. Then $\bar h_{\mu\nu}$ was determined in terms of a modified or effective matter-energy distribution. From this, the effective energy density was obtained as the usual energy density $T_{00}$, plus a perturbative correction proportional to $m^{-2}$, involving the Laplacian of the integral of $T$, weighted by the retarded propagator of the Klein-Gordon equation. Finally, we numerically computed the perturbative term in a binary star system, evaluating it as a function of $m$ and spatial position near the stars. In all cases, the results illustrate how the gravitational influence of the stars diminishes with distance. Additionally, the perturbation decreases as $m$ increases, consistently recovering the relativistic limit. These results highlight the role of modified gravity corrections in the vicinity of compact objects.

\end{abstract}

\maketitle


\section{\label{sec:level1}Introduction}
In recent decades, modified gravity theories have emerged as a significant alternative for explaining cosmological phenomena tied to fundamental physics \cite{Shankaranarayanan_2022,Clifton_2012}. These theories are motivated by the need to expand our understanding of gravity to address problems that General Relativity (GR) cannot solve without invoking dark matter and dark energy, such as the accelerated expansion of the universe \cite{Perlmutter_1997,Riess_1998} and the dynamics of large-scale structures \cite{Amendola_2007b,Starobinsky_2007}. Among these, $f(R)$ theories generalize the Einstein-Hilbert Lagrangian by introducing nonlinear terms in the Ricci scalar $R$, offering new perspectives on the curvature effects of spacetime across various gravitational regimes \cite{guth,Capozziello:2004km,Sotiriou_2006,Sotiriou:2008rp}. Viable models within $f(R)$ gravity theories \cite{Hough2020,ODINTSOV2023137988,Hurtado_2023}, show promise in addressing both cosmological and astrophysical phenomena beyond standard GR \cite{Nojiri,De_Felice_2010,Odintsov_2017,Chatterjee_2024}. These models introduce a broader range of gravitational behaviors that remain consistent with observations while avoiding many of the issues found in alternative theories \cite{Amendola_2007b,oikonomou,Sultana,SaezGomez2013}. Consequently, $f(R)$ gravity provides a compelling framework for exploring modifications to GR, with applications ranging from large-scale cosmic expansion to localized strong gravitational fields, such as black holes, where they predict modified horizon structures and unique stability properties \cite{Cruz,hurtadoRo,NASHED2021136133}. 
\\
In this sense, Starobinsky’s model \cite{Starobinsky:1980te}, $f(R)=R+\alpha R^2$, with $\alpha$ being a parameter; initially proposed within the framework of cosmic inflation, stands as one of the earliest alternatives to scalar-field inflation models \cite{Asaka_2016,ketov}. This model is characterized by the inclusion of an $R^2$ term, introducing new degrees of freedom in the field equations, resulting in additional modes that can impact the early Universe’s behavior \cite{Ivanov_2022, Asaka_2016}, particularly during inflation, where quantum effects may play a significant role \cite{Vilenkin,Giamalas}, as well as in producing perturbations that affect the evolution of large-scale structures such as galaxies and galaxy clusters.
\\
With the successful detection of gravitational waves \cite{Abbott2016}, the framework of $f(R)$ gravity could gain significant attention \cite{Sharif_2017,Gong_2018,Nakamura:2019Pf,Kalita_2021}, since important works have been developed in order to explore how these modified theories could influence wave propagation, polarization modes, and potential deviations from predictions in GR \cite{Vainio_2017,Liang2017,Katsuragawa_2019,Moretti2019,Gogoi_2021,ODINTSOV2022100950,Gegenberg_2018}. Hopefully the next-generation detectors will allow to observe a wider range of astrophysical events, providing further evidence of gravitational waves and their properties \cite{Caprini_2016,kalogera2021generationglobalgravitationalwave}. These advancements may reveal distinctive features in gravitational wave signals that could indicate the influence of $f(R)$ modifications, potentially setting them apart from standard relativistic predictions.
\\
Considering this context, we focus this work on linearizing the field equations of $f(R)$ gravity in the Starobinsky model within the weak-field approximation, to identify the distinctive deviations from GR.

This paper is organized as follows: in Section \ref{sec:level2}, we linearize the field equations of the $f(R)$ theory for the model $R+R^2/(6m^2)$ in the weak-field regime, expressing the field equations in terms of the perturbation $\bar h_{\mu\nu}$ relative to the background Minkowski metric. In Section \ref{sec:level3}, we solve the trace equation using Green's functions and auxiliary fields. In Section \ref{sec:level4}, we calculate the components of $\bar h_{\mu\nu}$, considering both the massive contribution induced by the quadratic term and the propagation effects of the field. Then, in Section \ref{sec:level5}, we numerically calculate the perturbation found for a binary star system. Finally, in Section \ref{sec:level6}, we discuss and analyze the results.

\section{\label{sec:level2}Field equations in $f(R)$ theory}
The $f(R)$ theory is an extension of GR, reformulating the Einstein-Hilbert action in terms of a nonlinear function of the Ricci curvature scalar $R$
\begin{equation}
    S=\frac{1}{2\kappa}\int f(R)\sqrt{-g}d^4x+S_M,
\end{equation}
where $S_M$ is the contribution from matter and energy, and $\kappa=8\pi$ ($c=G=1$). The field equations governing the dynamics of the metric tensor $g_{\mu\nu}$, relate the spacetime geometry to the distribution of matter and energy, and are derived by evaluating the critical points of the action $S$. In the metric formalism are written as
\begin{equation}\label{fieldequations}
    FR_{\mu\nu}-\frac{1}{2}fg_{\mu\nu}-F_{;\mu\nu}+g_{\mu\nu}F_{;\sigma}^{;\sigma}=\kappa T_{\mu\nu}, 
\end{equation}
where $f=f(R)$, $F=F(R)=f'(R)$, $F_{;\mu\nu}=\nabla_\mu\nabla_\nu F$ is the covariant derivative, and with the definition of the D'Alembertian operator $\Box=\nabla_\sigma\nabla^\sigma=g^{\sigma\rho}\nabla_\sigma\nabla_\rho$, $F_{;\sigma}^{;\sigma}=\Box F$. Moreover, the energy momentum tensor is defined as
\begin{equation}
    T_{\mu\nu}=\frac{-2}{\sqrt{-g}}\frac{\delta S_M}{\delta g^{\mu\nu}}.
\end{equation}
Equations (\ref{fieldequations}) incorporate the additional degrees of freedom introduced by the function $f(R)$, resulting in a framework where the curvature itself acts as a dynamical field, allowing the theory to be more flexible in addressing a broader range of gravitational phenomena. The trace of field equations (\ref{fieldequations}) is obtained by multiplying the equations by the metric tensor
\begin{equation}
    F R-2f+3F_{;\sigma}^{;\sigma}=\kappa T,
\end{equation}
where $T=T_{\mu\nu}g^{\mu\nu}$. Now, we will take into account the Starobinsky quadratic model \cite{Starobinsky:1980te}
\begin{equation}\label{starobinsky}
    f(R)=R+\frac{1}{6m^2}R^2,
\end{equation}
where the parameter $m$ is identified with the inflaton mass \cite{Asaka_2016,ketov}. With this function of $f(R)$, field equations take the form
\begin{equation}\label{fieldeq2}
    G_{\mu\nu}-\frac{1}{3m^2}\left[\left(\frac{1}{4}R^2+R_{;\sigma}^{;\sigma}\right)g_{\mu\nu}+RR_{\mu\nu}-R_{;\mu\nu}\right]=\kappa T_{\mu\nu},
\end{equation}
where we have used the Einstein tensor
\begin{equation}
    G_{\mu\nu}=R_{\mu\nu}-\frac{1}{2}Rg_{\mu\nu}.
\end{equation}
Considering minimal deviations, $h_{\mu\nu}$, from flat spacetime, that is, small fluctuations around the background Minkowski metric, $\eta_{\mu\nu}=\text{diag}(-1,1,1,1)$, so that the metric tensor can be expressed as
\begin{equation}
    g_{\mu\nu}=\eta_{\mu\nu}+h_{\mu\nu},
\end{equation}
such that $h_{\mu\nu}\ll 1$. In this weak-field approximation, the quadratic terms are not taking into account in equations (\ref{fieldeq2}), therefore they are simplified to
\begin{equation}\label{fielequationssimple}
    G_{\mu\nu}+\frac{1}{3}\left(\eta_{\mu\nu}R_{;\sigma}^{;\sigma}-R_{;\mu\nu}\right)=\kappa T_{\mu\nu},
\end{equation}
and using the definition of the trace reverse tensor 
\begin{equation}
    \bar h_{\mu\nu}=h_{\mu\nu}-\frac{1}{2}\eta_{\mu\nu}h,
\end{equation}
through which it is possible to simplify the Einstein tensor, that is
\begin{equation}
    G_{\mu\nu}=-\frac{1}{2}\bar h_{\mu\nu,\sigma}^{,\sigma},
\end{equation}
and by choosing the Lorentz gauge condition $\bar h^{\mu\nu}_{,\nu}=0$, we arrive to the linearized field equations
\begin{equation}\label{linearized}
    \bar h_{\mu\nu,\sigma}^{,\sigma}-\frac{1}{3m^2}\left(\eta_{\mu\nu}\bar h_{,\sigma\rho}^{,\sigma\rho}-\bar h_{,\sigma\mu\nu}^{,\sigma}\right)=-2\kappa T_{\mu\nu},
\end{equation}
that is, in the formulation of $f(R)$ gravity using the Starobinsky model, the additional term $1/(3m^2)$ acts as a correction that introduces mass effects associated with the extra degrees of freedom, particularly with the scalar field incorporated through the $R^2$ term. This expression is more complex than in GR, and its deviation reflects the additional modes due to the effective mass $m$. However, in the limit $m\to\infty$ (so that $f(R)\to R$ in Eq. (\ref{starobinsky})), the additional term vanishes, and the standard wave equation of GR is recovered.
Taking the trace of Eq. (\ref{linearized}) we obtain
\begin{equation}
    \frac{1}{m^2}\bar h_{,\sigma\rho}^{,\sigma\rho}-\bar h_{,\sigma}^{,\sigma}=2\kappa T,    
\end{equation}
or using the D'Alembertian operator in flat space $\Box=-\partial_t^2+\nabla^2$,
\begin{equation}\label{trace}
    \left(\Box-m^2\right)\Box \bar h(x^\sigma)=2\kappa m^2T(x^\sigma),
\end{equation}
thus, field equations (\ref{linearized}) are rewritten as
\begin{equation}
    \Box\left(3\bar h_{\mu\nu}-\eta_{\mu\nu}\bar h+\frac{1}{m^2}\bar h_{,\mu\nu}\right)=-2\kappa\left(3T_{\mu\nu}-T\eta_{\mu\nu}\right),
\end{equation}
and returning to the initial field $h$
\begin{equation}
    \Box\left(6 h_{\mu\nu}-\eta_{\mu\nu}h-\frac{2}{m^2}h_{,\mu\nu}\right)=-4\kappa\left(3T_{\mu\nu}-T\eta_{\mu\nu}\right).
\end{equation}
\section{\label{sec:level3}Solution of the trace equation}
Equation (\ref{trace}) describes the evolution of a field $\bar h(x^\sigma)$ under the influence of a source $T(x^\sigma)$, with a mass term $m^2$. Due to its structure, it can be divided into two stages, each with a distinct physical meaning and mediated by a fictitious field. To see this, suppose $H(x^\sigma)$ is a fictitious field generated by the real source $T(x^\sigma)$; it can be regarded as a perturbation or an ``auxiliary wave'' that propagates as a massless field:
\begin{equation}\label{HT}
    \Box H(x^{\sigma})=-2\kappa T(x^{\sigma}),
\end{equation}
thus, by Eq. (\ref{trace}), we can think of $H(x^{\sigma})$ as a fictional source that produces $\bar h(x^\sigma)$ through the inhomogeneous Klein-Gordon equation
\begin{equation}\label{hH}
    \left(\Box-m^2\right)\bar h(x^\sigma)=-m^2 H(x^\sigma),
\end{equation}
where we see that the propagation of the perturbation is modified by the presence of the mass term. Interpreting Eq (\ref{trace}) in two stages, reflects the fact that modified gravity can introduce interactions or auxiliary fields that mediate the effective gravitational response.
\\
To solve these coupled differential equations, Green's functions can be used, as they hold significant physical meaning regarding the propagation of perturbations in spacetime. Suppose $G_1$ is the solution of
\begin{equation}\label{eqgreen1}
    \Box_xG_1(x^\sigma-z^\sigma)=\delta^{(4)}(x^\sigma-z^\sigma),
\end{equation}
where $\Box_x$ denotes derivative with respect coordinates $x^{\sigma}$ and $\delta^{(4)}(x^\sigma-z^\sigma)$ is the four dimensional Dirac delta function. Therefore, we can express $\bar H(x^{\sigma})$ as the sum of all signals to the past generated by the source $T$ in the point $z^{\sigma}$, i.e.
\begin{equation}\label{HTintegral}
    H(x^{\sigma})=-2\kappa \int T(z^{\sigma})G_1\left(x^{\sigma}-z^{\sigma}\right)d^4z,
\end{equation}
and equivalently, if $G_2(x^{\sigma}-y^{\sigma})$ satisfies
\begin{equation}\label{eqgreen2}
    \left(\Box_x-m^2\right)G_2(x^\sigma-y^\sigma)=m^2 \delta^{(4)}(x^\sigma-y^\sigma),
\end{equation}
we can express the field $\bar h$ at the point $x^\sigma$, as
\begin{equation}\label{hHintegral}
    \bar h(x^\sigma)=-\int H(y^\sigma)G_2(x^\sigma-y^\sigma)d^4y,
\end{equation}
or in terms of the source $T$,
\begin{equation}\label{solh1}
    \bar h(x^{\sigma})=2\kappa\int\int T(z^{\sigma})G_2(x^{\sigma}-y^{\sigma})G_1(y^{\sigma}-z^{\sigma})d^4yd^4z,
\end{equation}
in other words, the function $G_1$ represents the spacetime response to the source $T$ located at the point $z^\sigma$. It describes the propagation of the perturbation from $z^\sigma$ to the point $y^\sigma$, generating the field $H$ there. In turn, $G_2$ propagates the influence of $H$ both on and inside the light cone (at subluminal speeds) to the point $x^\sigma$, where the field $\bar h$ is measured.
\\
Note that at the limit $m\to\infty$ in Eq. (\ref{hH}), $\bar h(x^\sigma)=H(x^{\sigma})$, and Eq. (\ref{HT}) takes the same form as in the GR case. Equivalently,
\begin{equation}
    \frac{1}{m^2}\Box\bar h \overset{m\to\infty}{\longrightarrow} 0, 
\end{equation}
by virtue of the finiteness of $T$. From a physical perspective, in this limit, the propagation of the field becomes constrained, and the Green’s function is concentrated in an infinitesimally small region around $y^\sigma$. Mathematically, this is expressed as $G_2(x^\sigma-y^\sigma)\to-\delta^{(4)}(x^\sigma-y^\sigma)$, so that the influence of the massive field is localized at the point $y^{\sigma}$ and rapidly diminishes outside of it.
\\
The solution of Eq. (\ref{eqgreen1}) is the well known retarded Green function
\begin{equation}\label{G1}
    G_1(x^\sigma-z^\sigma)=-\frac{\delta(s_{xz}^2)}{2\pi}\theta(t_{xz}),
\end{equation}
where the theta function, $\theta(x)=1$ only for $x>0$, $s_{xz}^2=t_{xz}^2-r_{xz}^2$, $t_{xz}=x^0-z^0$ and $r_{xz}^2=\sum_{i=1}^3(x^i-z^i)^2$. To ensure the propagator is non-zero in Minkowski spacetime, the signal generated at $z^{\sigma}$ must lie in the past of the event $x^{\sigma}$, with a positive proper time interval between them, thereby upholding causality. Likewise, the solution for Eq. (\ref{eqgreen2}) is
\begin{equation}\label{G2}
    G_2(x^\sigma-y^\sigma)=m^2\theta(t_{xy})\left[-\frac{\delta(s_{xy}^2)}{2\pi}+g(x^\sigma-y^{\sigma})\right],
\end{equation}
where
\begin{equation}
    g(x^\sigma-y^\sigma)=\frac{m}{4\pi}\theta(s_{xy}^2)\frac{J_1(m s_{xy})}{s_{xy}},
\end{equation}
$J_1(x)$ is the Bessel function of the first kind, and $x^0>y^0>z^0$, from which 
\begin{equation}
    \delta(s_{xz}^2)=\frac{\delta(t_{xz}-r_{xz})}{2 r_{xz}}.
\end{equation}
It is worth noting here that Eq. (\ref{trace}) can be decomposed from a different perspective: suppose the field $\bar h(x^\sigma)$ is generated by a fictitious source $K(x^\sigma)$ through the wave equation
\begin{equation}\label{hI}
\Box \bar{h}(x^{\sigma}) = -K(x^{\sigma}),
\end{equation}
where the source $K$ is produced by the real source $T(x^\sigma)$ according to
\begin{equation}\label{IT}
\left(\Box - m^2\right) K(x^{\sigma}) = -2\kappa m^2 T(x^{\sigma}).
\end{equation}
which, again, is the inhomogeneous Klein-Gordon equation. However, when expressing $T$ in integral form using the delta function
\begin{equation}
    T(x^{\sigma})=\Box_x\int T(z^{\sigma})G_{1}(x^{\sigma}-z^{\sigma})d^4z,
\end{equation}
and for Eq. (\ref{HTintegral}), we found the relation between the fields $H$ and $K$,
\begin{equation}
    \left(\Box-m^2\right)K(x^{\sigma})=m^2\Box H(x^{\sigma}),
\end{equation}
this equation indicates that the two fictitious fields are not independent; rather, the physical field $\bar h$ produced by the real source $T$ can be interpreted as being mediated by the coupled fields $K$ and $H$. In this scheme, $T$ generates $H$, $H$ generates $K$, and $K$ in turn produces $\bar h$.
\\
Eq. (\ref{IT}) can be written as
\begin{equation}
    K(x^\sigma)=-2\kappa\int T(z^\sigma)G_2(x^\sigma-z^\sigma)d^4z,
\end{equation}
and from Eq. (\ref{hI}),
\begin{equation}\label{solh2}
    \bar h(x^{\sigma})=2\kappa\int\int T(z^{\sigma})G_1(x^{\sigma}-y^{\sigma})G_2(y^{\sigma}-z^{\sigma})d^4yd^4z,
\end{equation}
since the physics described by the physical field $\bar h$ must be invariant in both approaches, the last integral must be equivalent to (\ref{solh1}). Therefore, the integrands can be equated
\begin{equation}\label{symmetry}
    G_1(x^{\sigma}-y^{\sigma})G_2(y^{\sigma}-z^{\sigma})=G_2(x^{\sigma}-y^{\sigma})G_1(y^{\sigma}-z^{\sigma}),
\end{equation}
which can be showed from the definitions of the Green functions. This equality indicates a commutation symmetry between the Green’s functions $G_1$ and $G_2$, suggesting that when applied sequentially for field propagation, the order in which they are used does not affect the final result. With this, we arrive to the important relation
\begin{multline}\label{integral}
    \int G_1(x^{\sigma}-y^{\sigma})G_2(y^{\sigma}-z^{\sigma})d^4y=\\
    \frac{1}{m^2}G_ 2\left(x^{\sigma}-z^{\sigma}\right)-G_ 1\left(x^{\sigma}-z^{\sigma}\right).
\end{multline}
this expression describes how the propagation of signals from point $z^\sigma$ to $x^{\sigma}$ depends on a linear combination of the functions $G_1$ and $G_2$, where the mass $m$ controls the types propagation. It shows that the propagation of the perturbation includes the influence of both a massless and a massive field. Note that we could also have arrived at this integral from Eq. (\ref{eqgreen2}),
\begin{equation}
    G_ 2 \left(x^{\sigma} - z^{\sigma}\right)=\Box \left[\frac{1}{m^2}G_2\left(x^{\sigma}-z^{\sigma}\right)-G_1\left(x^{\sigma}-z^{\sigma}\right)\right].
\end{equation}
With Eq. (\ref{integral}) we can write the trace Eq. (\ref{solh2}) as
\begin{multline}
    \bar h(x^{\sigma})=\frac{\kappa}{2\pi}\int\frac{1}{r_{xz}}T\left(x^0-r_{xz},z^i\right)d^3z\\ +\frac{2\kappa}{m^2}\int T\left(z^{\sigma}\right)G_2\left(x^{\sigma}-z^{\sigma}\right)d^4z,
\end{multline}
where it can be seen the first term that corresponds to GR, and a second term that disappears when $m\to \infty$. However, the trace can be written in a simpler way by using the bulk term of $G_2$, since
\begin{equation}
    \int G_1(x^{\sigma}-y^{\sigma})G_2(y^{\sigma}-z^{\sigma})d^4y=g_ 2\left(x^{\sigma}-z^{\sigma}\right),
\end{equation}
therefore
\begin{equation}\label{tracesolution}
\begin{split}
    \bar h(x^{\sigma})&=2\kappa\int T(z^{\sigma})g(x^{\sigma}-z^{\sigma})d^4z\\
    &=\frac{\kappa m} {2\pi}\int_{-\infty}^{x^0-r_{xz}} T\left(z^{\sigma}\right)\frac{J_ 1\left(m s_{xz}\right)}{s_{xz}}d^4z,
\end{split}
\end{equation}
note that the integrand has an oscillatory behavior due to the Bessel function, whose frequency increases with $m$, as does its amplitude, however as the interval $s_{xz}$ increases, the intensity of the perturbation decreases.

\section{\label{sec:level4}Solution of the field equations}
Now that we have established the dependence of $\bar h(x^\sigma)$ on $T(x^{\sigma})$ and the function $g(x^\sigma-z^\sigma)$, we can solve Eq. (\ref{linearized}) for the field $\bar h_{\mu\nu}$, as a combination of derivatives and integrals of the stress-energy tensor, along with the Green’s functions. In this sense, note that
\begin{equation}
    \Box g(x^\sigma-z^\sigma)=G_2(x^\sigma-z^\sigma),
\end{equation}
so that Eq. (\ref{tracesolution}) takes the form
\begin{equation}
    \Box \bar {h}(x^{\sigma})=2\kappa\int T\left(z^{\sigma}\right) G_2\left(x^{\sigma}-z^{\sigma}\right)d^4z,
\end{equation}
and
\begin{equation}
    \Box^2 \bar {h}(x^{\sigma})=2\kappa m^2\left(T\left(x^{\sigma}\right)+\int T\left(z^{\sigma}\right)G_2\left(x^{\sigma}-z^{\sigma}\right)d^4z\right),
\end{equation}
by substituting these two results into Eq. (\ref{linearized}), we obtain a wave equation for the tensor $\bar{h}_{\mu\nu}$
\begin{equation}\label{wavemodi}
    \square\bar{h}_{\mu\nu}\left(x^{\sigma}\right)=-2\kappa\mathcal{T}_{\mu\nu}(x^\sigma),
\end{equation}
where the derivatives are respect to $x^{\sigma}$. Therefore we can interpret that the perturbation is produced by a modified, or effective, stress-energy tensor
\begin{widetext}
\begin{equation}\label{tmodif}
    \mathcal{T}_{\mu\nu}(x^\sigma)=T_{\mu\nu}\left(x^{\sigma}\right)-\frac{1}{3}\eta_{\mu\nu}T\left(x^{\sigma}\right)-\frac{1}{3}\left(n_{\mu\nu}-\frac{1}{m^2}\partial_{\mu}\partial_{\nu}\right)\int T\left(z^{\sigma}\right) G_2\left(x^{\sigma}-z^{\sigma}\right)d^4z.
\end{equation}
Using the Green's function $G_1$ again, we can write the solution to the differential equation (\ref{wavemodi}), that is
\begin{equation}\label{tensorpert}
    \bar{h}_{\mu \nu }\left(x^{\sigma }\right)=\frac{2 \kappa}{3}\int \left[\frac{\eta_{\mu \nu}}{m^2}T\left(y^{\sigma }\right) G_2\left(x^{\sigma }-y^{\sigma }\right)-\left(3 T_{\mu \nu }\left(y^{\sigma }\right)+\frac{1}{m^2}\int T\left(z^{\sigma }\right)\partial_\mu\partial_\nu G_2\left(y^{\sigma }-z^{\sigma }\right)d^4z\right)G_1(x^\sigma-y^\sigma)\right]d^4y,
\end{equation}
\end{widetext}
where it should be noted that the derivatives are now with respect to $y^\sigma$. Likewise, it is observed that at $m\to\infty$, the terms associated with the trace $T$ disappear, and the case of GR is obtained.
\\
Since $\mathcal{T}_{00}$ incorporates the effects of the modified theory, including the mass terms, we can express the quadrupole moment tensor as
\begin{equation}\label{quadrupole}
I_{ij}(t) = \int z_i z_j \, \mathcal{T}_{00}(t, z_k) \, d^3z.
\end{equation}
This approach is advantageous because it allows us to integrate directly over the effective energy density, revealing how the source responds under the influence of the additional terms. Therefore, from (\ref{tmodif})
\begin{multline}
    \mathcal{T}_{00}\left(x^{\sigma }\right)=T_{00}\left(x^{\sigma }\right)+\frac{1}{3}T\left(x^{\sigma }\right)+\frac{1}{3}\left(1+\frac{1}{m^2}\partial _{0}^2\right)\\\int T\left(z^{\sigma }\right)G_2\left(x^{\sigma }-z^{\sigma }\right)d^4z,  
\end{multline}
or equivalently
\begin{equation}\label{t00}
    \mathcal{T}_{00}(x^\sigma)=T_{00}(x^\sigma)+\frac{1}{3 m^2}\nabla^2\int T\left(z^{\sigma}\right) G_2\left(x^{\sigma}-z^{\sigma}\right)d^4z,
\end{equation}
this expression shows how the effective energy density is the sum of the direct component of the source, $T_{00}$, plus a correction that fades as $m$ increases, but is responsible for the finite-range effects. In the same way the energy density can be expressed as follows
\begin{equation}\label{source}
    \mathcal{T}_{00}\left(x^{\sigma }\right)=T_{00}(x^\sigma)-\frac{1}{3 m^2}\partial_0^2T(x^\sigma)+\frac{1}{3 m^4}\mathfrak {T},
\end{equation}
where
\begin{equation}\label{termcorrection}
    \mathfrak {T}=\Box_x\left(m^2+\partial_0^2\right)\int T\left(z^{\sigma}\right)G_2\left(x^{\sigma}-z^{\sigma}\right)d^4z.
\end{equation}
This separation highlights how the contributions from the modified theory decay with $m$. In particular, the term $\partial_0^2T(x^\sigma)/(3 m^2)$ captures dynamical effects for finite $m$ without requiring integration, offering a practical approach to analyze the system.

Finally, the spacelike components of the perturbation can be written as
\begin{equation}\label{finalpert}
    \bar h_{ij}(t,x)=\frac{2}{r}\partial_0^2I_{ij}(t_r),
\end{equation}
where the retarded time is defined as $t_r=t-r$, that is, the time in which the wave is observed after being generated by the system. 
\section{\label{sec:level5}Binary star system}
Due to the highly oscillatory nature of the integrand in $\mathcal{T}_{00}(x^\sigma)$, particularly the term involving the Bessel function in $G_2\left(x^{\sigma}-z^{\sigma}\right)$, significant challenges arose in computing these terms directly. As an illustrative example, let us consider a binary star system, as described in \cite{carroll2019}. While this system is a natural candidate for exploring the components of the quadrupole moment tensor (\ref{quadrupole}) and the perturbation term (\ref{finalpert}) in the modified theory, the analytical integration of $ T\left(z^{\sigma}\right)G_2\left(x^{\sigma}-z^{\sigma}\right)$ proved to be intractable. Nevertheless, valuable information about the system can be extracted by numerically computing the correction term in Eq. (\ref{t00}), that is
\begin{equation}\label{correction}
    \frac{1}{3 m^2}\nabla^2\int T\left(z^{\sigma}\right) G_2\left(x^{\sigma}-z^{\sigma}\right)d^4z,
\end{equation}
at specific spatial locations. This approach allowed us to observe the behavior of the perturbation term as a function of $m$ and validate its expected decay for large $m$, consistent with the recovery of General Relativity in this limit.
\\
Let the positions of two stars of mass $M$, at a distance $r$ from the point of observation, orbiting each other due to their mutual gravitational attraction, at a distance $R$ from the system's center of mass, and with orbital angular frequency $\Omega$, be
\begin{equation}
    x_a^1(t) = R\cos(\Omega t), \quad x_a^2(t) = R\sin(\Omega t),
\end{equation}
and
\begin{equation}
    x_b^1(t) = -R\cos(\Omega t), \quad x_b^2(t) =-R\sin(\Omega t),
\end{equation}
from which
\begin{subequations}\label{tensorT}
	\begin{align}
		T_{00} (t, x^i) &= M\left(\delta_a+\delta_b\right),   \label{T1} \\
		T_{11} (t, x^i)&= M  \Omega^2R^2\sin^2(\Omega t)\left(\delta_a+\delta_b\right),\label{T2} \\
		T_{22} (t, x^i)&= M \Omega^2R^2\cos^2(\Omega t)\left(\delta_a+\delta_b\right),  \label{T3}\\
        T_{33} (t, x^i)&= 0,  \label{T4}
	\end{align}
\end{subequations}
and the trace
\begin{equation}
    T(t,x^i)=M\left(\Omega^2R^2 -1\right)\delta\left(x^3\right)\left(\delta_a+\delta_b\right),
\end{equation}
where
\begin{equation}
     \delta_{a \above 0pt b}=\delta\left[x^1\mp R\cos(\Omega t)\right]\delta\left[x^2\mp R\sin(\Omega t)\right].
\end{equation}
For this system the components of the quadrupole moment tensor can be written as
\begin{equation}
    I_{11} = M R^2  (1+ \xi\cos  (2  \Omega t))+\mathfrak {I}_{11},
\end{equation}
\begin{equation}
    I_{22} = M R^2  (1- \xi\cos  (2  \Omega t))+\mathfrak {I}_{22},
\end{equation}
\begin{equation}
    I_{12} = I_{21} = M R^2  \xi\sin  (2  \Omega t)+\mathfrak {I}_{12},
\end{equation}
where 
\begin{equation}
    \xi=1-\frac{4\Omega^2}{3m^2}+\frac{4R^2\Omega^4}{3m^2},
\end{equation}
and $\mathfrak {I}$ represents the contribution generated by term (\ref{termcorrection}), which cannot be calculated analytically.
Therefore, it is observed that the factor $\xi$ introduces modifications to the amplitude of the perturbation due to the mass of the field. When $m \to \infty$, $\mathfrak{T}\to 0$ and $\xi \to 1$, recovering the relativistic result. However, from a classical perspective
\begin{equation}
    \Omega^2 = \frac{M}{4R^3},
\end{equation}
and 
\begin{equation}
    \xi=1- \frac{M}{3 m^2 R^3}+\frac{M^2}{12 m^2R^4},
\end{equation}
remarkably, for $R = M/4$, we find $\xi = 1$. In this case, the modified theory's contribution to the system dynamics would be entirely determined by the term $\mathfrak{T}$.



\begin{figure}[h]
    \centering
    \includegraphics[width=0.5\textwidth]{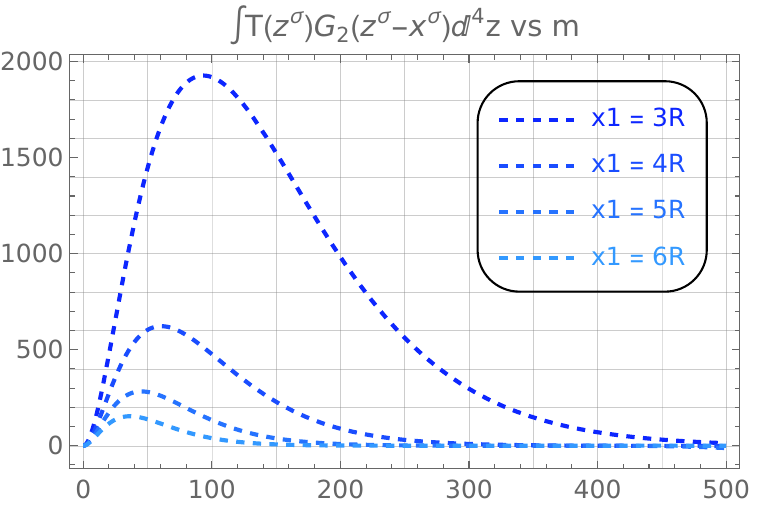}
    \caption{Numerical evaluation of $\int T\left(z^{\sigma}\right) G_2\left(x^{\sigma}-z^{\sigma}\right)d^4z$, in $L^{-2}$, as a function of $m$, in $L^{-1}$, for different values of $x^1$ and $x^0=x^2=x^3=0$. The curves, from top to bottom, correspond to $x^1=3R, 4R, 5R$, and $6R$. As $x^1$ increases, the amplitude of the integral decreases, and the curves decay to zero more rapidly (i.e., at lower values of $m$).}
    \label{fig:1}
\end{figure}

Although the term $\mathfrak{T}$ could not be computed analytically, all relevant information is encoded in $\int T\left(z^{\sigma}\right) G_2\left(x^{\sigma}-z^{\sigma}\right)d^4z$. To gain insight into its behavior, we performed a numerical evaluation, setting $M=1$, $R=0.01$, $\Omega=2\pi$ (in geometrized units $[M]=[R]=L$, $[\Omega]=L^{-1}$), and focusing on spatial points near one of the stars. The results, shown in Fig. \ref{fig:1}, reveal a functional dependence of the form $b m^2e^{-a m^n}$, where $a$, $b$, and $n$ are parameters that depend on the coordinate $x^1$. This exponential decay with $m$ confirms the suppression of non-GR effects at high mass scales, consistent with the expected recovery of General Relativity in the $m\to\infty$ limit.

\begin{figure}[h]
    \centering
    \includegraphics[width=0.49\textwidth]{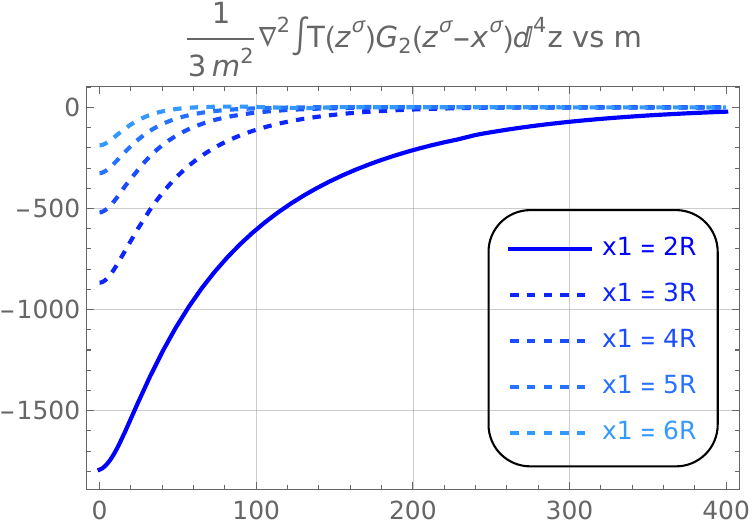}
    \caption{Correction term, Eq. (\ref{correction}), in units of $L^{-2}$ as a function of $m$ for different values of $x^1$ (bottom to top: $2R$ [continuous line], $3R$, $4R$, $5R$, $6R$), obtained using a global adaptive integration method, and with $x^0=x^2=x^3=0$. Each curve starts at a negative value and increases smoothly toward zero, exhibiting a concave-down shape. As $x^1$ increases, the amplitude of the correction term (i.e., its value at $m\to0$) decreases, tending to zero.}
    \label{fig:2}
\end{figure}

\begin{figure}[h]
    \centering
    \includegraphics[width=0.49\textwidth]{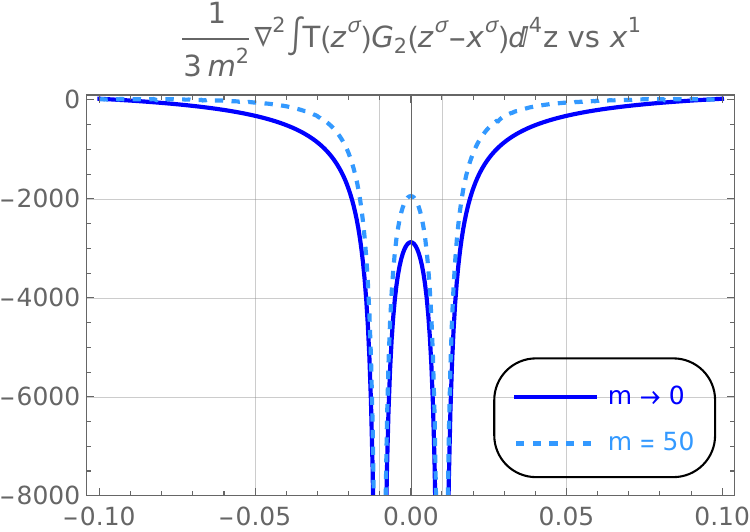}
    \caption{Spatial dependence of the correction term as a function of $x^1$, in units of $L$, for $m=0$ and $m=50$, and for $x^0=x^2=x^3=0$. The curves reflect the suppression of modified gravity effects at larger distances. The units of the vertical axis are $L^{-2}$.}
    \label{fig:3}
\end{figure}

\begin{figure}[h]
    \centering
    \includegraphics[width=0.47\textwidth]{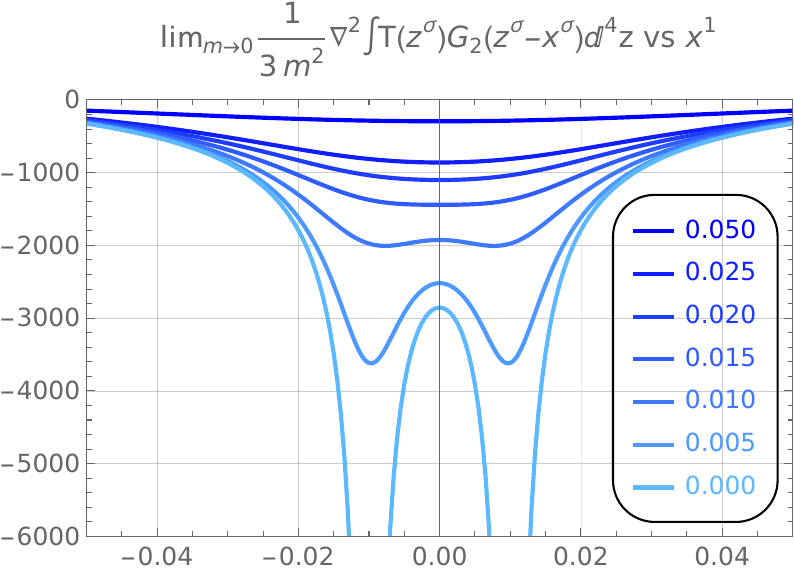}
    \caption{Perturbation ($L^{-2}$) as a function of $x^1$ ($L$) for different values of $x^3$ (shown in the inset box, corresponding to each curve from bottom to top), plotted in the limit $m\to 0$, and for $x^0=x^2=0$. All curves tend to zero, consistent with the expected suppression of perturbations far from the source.}
    \label{fig:4}
\end{figure}

\noindent We also calculated the correction term, Eq. (\ref{correction}), for similar values of $x_1$ and as a function of $m$. The results are shown in Fig. \ref{fig:2}. The observed value of each curve as $m$ tends to zero can be explained by the dominance of the delta function term in $G_2$, while the contribution from the Bessel function becomes negligible. The delta function term is negative, and this sign is preserved by the action of the Laplacian. The $m^{-2}$ factor amplifies the magnitude of the term, but its divergence is regulated by the combined effect of the integral and the Laplacian.
In addition, we computed the correction term as a function of the spatial variable $x^1$, as illustrated in Fig. \ref{fig:3}. This figure complements the previous one by showing the spatial dependence of the modified gravity effects. Near the stars, located at $x^1=R$ and $x^1=-R$, the correction term exhibits a sharp decrease, reflecting the intense gravitational influence of the stars. As $x^1$ moves away from the stars, the perturbation gradually diminishes, eventually approaching zero. This behavior is consistent with the expected decay of modified gravity effects at larger distances.
\\
Finally, Fig. \ref{fig:4} shows the perturbation as a function of $x^1$, this time for different values of $x^3$ (i.e., as the perpendicular distance from the center of mass of the system increases). The curves are plotted for $m\to 0$, highlighting the behavior of the perturbation in the limit where modified gravity corrections are most significant.




\section{\label{sec:level6}Conclusions}
This work explores the linearized field equations in the Starobinsky $R+R^2/(6m^2)$ model, deriving an effective matter-energy distribution for perturbations through an auxiliary field, $H(x^\sigma)$, that links the trace equation (\ref{trace}), the Klein-Gordon equation (\ref{hH}), and wave equation (\ref{HT}). This two-step propagation reflects how $H(x^\sigma)$ mediate the gravitational effect in an indirect way.
Equations (\ref{HT}) and (\ref{hH}) were solved using Green’s functions, (\ref{G1}) and (\ref{G2}). It should be noted that $G_1(x^\sigma-z^\sigma)$ models the direct propagation of the perturbation at light speed from the source, while $G_2(x^\sigma-z^\sigma)$ allows for the additional propagation of the field $H(x^\sigma)$ with massive characteristics, creating both light-speed and slower components in the gravitational response measured at $x^\sigma$.
For this massive field, the Green’s function, $G_2$ shows exponential decay: the larger the mass, the more localized the field's effect near the source, in a similar way to the Yukawa potential: a massive field’s influence decays with distance, reflecting a finite \textit{interaction range}. In the limit $m\to \infty$, $G_2$ approaches a delta-like behavior, confining the field entirely to the source point, in contrast to the infinite range of a massless field.
\\
We found the trace $\bar h(x^\sigma)$, Eq. (\ref{solh2}), which could be simplified due to Eq. (\ref{symmetry}). This symmetry can be understood in terms of the commutativity in combining these solutions for the propagation of perturbations between the points $z^\sigma\to y^{\sigma}\to x^\sigma$. This property could be useful in coupled field theories, where propagation and interaction between points in spacetime can be decomposed into multiple steps.
\\
The solution found for the trace $\bar h(x^\sigma)$, Eq. (\ref{tracesolution}), is written in terms of the of the Bessel function, suggesting that the propagation of $\bar h$ exhibits an oscillatory and finite-range nature. This is because the mass term restricts the propagation to a scale defined by $1/m$. At greater distances, the Bessel function decays, reflecting the effect of the mass in limiting the influence of the source. Additionally, the term $J_1(ms_{xz})/s_{xz}$ has a decay controlled by $s_{xz}$, indicating that the intensity of the perturbation decreases with the distance between $x^{\sigma}$ and $z^\sigma$. As the mass increases, the term $J_1(ms_{xz})/s_{xz}$ narrows and concentrates near $s_{xz}=0$, while its amplitude asymptotically grows toward infinity, exhibiting a delta-like behavior in the limit $m\to\infty$. This transition of the function $g(x^\sigma)$ toward $G_1(x^\sigma)$ in this limit reflects that, as the massive field becomes extremely heavy, its influence on the effective field becomes confined to the source point.
\\
Once the trace was replaced in the field equations, it could be observed that the field $\bar h(x^\sigma)$ was defined by a wave equation, (\ref{wavemodi}), with an effective stress-energy tensor, Eq. (\ref{tmodif}). This modified source term incorporates contributions not only from the actual energy-momentum tensor $T_{\mu\nu}$ but also from additional terms mediated by the Green’s functions and differential operators, which capture the effects of the massive field components on the propagation of $\bar{h}_{\mu\nu}$. This effective source modifies the propagation characteristics of the wave equation, reflecting the impact of mass terms and other higher-order effects on the field.
\\
We found the solution for the tensor $\bar h_{\mu\nu}$, Eq. (\ref{tensorpert}). The terms within the integral, represent both massless and massive field contributions: $T_{\mu\nu}$ corresponds to the massless component, while the terms associated with the trace $T$ are modulated by the mass $m$. Consequently, the effect of $T$ on $\bar h_{\mu\nu}$, can reach over long distances for small masses or becoming locally confined for large masses. In fact, as $m\to\infty$, the trace terms vanish, recovering the case of GR.
\\
We have expressed the energy density, Eq. (\ref{t00}), as the sum of the $T_{00}$ component and a correction term, Eq. (\ref{correction}). This correction term captures the additional contributions arising from the modified gravity theory, inversely proportional to $m^2$, a behavior that is consistent with the expected suppression of perturbations with the increase of $m$, validating the physical interpretation of the $f(R)$ corrections.
\\
A binary system model was used to numerically compute the perturbation as a function of the parameter $m$ and the coordinate $x^1$ at specific spatial points: near one of the stars and at various distances from the center of mass of the system. The numerical approach was adopted because the oscillatory nature of the Bessel function made an analytical calculation of the perturbative term impractical for this system.
\\
In all cases studied, the perturbative term decays asymptotically to zero as $m$ increases. This behavior aligns with the expected recovery of General Relativity in the limit $m\to\infty$, as the additional contributions from the modified theory vanish, confirming the consistency of our results with the theoretical predictions.
\nocite{*}
\bibliography{apssamp}

\end{document}